\documentclass[aps,prd,preprint,eqsecnum,nofootinbib,superscriptaddress]{revtex4}
\usepackage{amssymb,amsmath,psfrag,epsfig}

\begin{document}

\noindent \hfill Brown-HET-1579

\vskip 1 cm

\title{Correlation Functions in Non-Relativistic Holography }

\author{Anastasia Volovich}

\affiliation{Brown University, Providence, Rhode Island 02912, USA}

\author{Congkao Wen}

\affiliation{Brown University, Providence, Rhode Island 02912, USA}

\begin{abstract}

Recently constructed gravity solutions with Schr\"odinger
symmetry provide a new example of AdS/CFT-type dualities for
the type of non-relativistic field theories relevant to certain
cold atom systems.
In this paper we use the gravity side to
calculate $n$-point 
correlation functions of scalar fields by reducing the
computation to that in ordinary AdS space via a particular
Fourier transform.
We evaluate the relevant integrals for 3- and 4-point
functions and show that the results are consistent with the
requirements of Schr\"odinger invariance, the implications of which
we also work out for general $n$-point functions.
\end{abstract}

\maketitle

\section{Introduction}

Gauge/string duality is a central theme of modern research
in string theory \cite{adscft,witten:1998,gkp}.  
As a strong/weak coupling duality it provides us
with a wealth of data useful
for studying interesting field theories at strong coupling
where other methods fail.
Its
applications currently range 
from nuclear physics to
plasma physics to tabletop
condensed matter systems \cite{applications}.

Until recently the study of gauge/gravity duality has been limited
to relativistic field theories. However many non-relativistic field
theories clearly play important roles in various physical systems.
An example of a system with non-relativistic conformal symmetry is a system of
fermions at unitarity which can be realized experimentally in certain cold
atom systems~\cite{experimetal}. This theory is invariant under a
non-relativistic conformal group called the Schr\"odinger group, and
it is natural to wonder if there exists a gravity dual which would
allow us to use the AdS/CFT correspondence 
to learn more about these systems.

Recently a new exciting example of such duality has
emerged. The simplest geometries with Schr\"odinger group isometry
have been constructed in~\cite{D.T.Son:2008,McGreevy:2008}, leading
to what is sometimes referred to as the AdS/cold atom
correspondence or non-relativistic AdS/CFT. 
Excitingly, these geometries have been embedded in
string theory in~\cite{Maldacena, Adams, Herzog}. These papers have
opened up a new avenue for studying non-relativistic conformal field
theories (NRCFTs)
at strong coupling. See~\cite{NRcft} for recent studies of various
aspects of non-relativistic AdS/CFT.

The holographic dictionary has been developed
in~\cite{D.T.Son:2008,McGreevy:2008} and the two-point function of
a scalar field was recovered from a gravity calculation
in~\cite{D.T.Son:2008,McGreevy:2008}. Unlike more familiar
applications of AdS/CFT where the boundary is one
dimension less than the bulk, here the boundary theory has two fewer
dimensions than the bulk~\cite{D.T.Son:2008,McGreevy:2008}.

In this paper we consider higher point correlation functions in the
gravity theory. We employ the trick introduced in~\cite{Henkel:2003}
where the free massive Schr\"odinger equation was reduced to a
non-massive Klein-Gordon equation via Fourier transform with respect
to the mass. Applying the same trick to the solution of the scalar
wave equation in the gravity background for non-relativistic field theory,
we reduce the wave
equation to that in ordinary AdS space. The computation of
correlation functions in the bulk then reduces to performing a
particular Fourier transform of AdS correlators. 
We explicitly evaluate the relevant
integrals for three- and four-point functions and show that the results are
indeed consistent with the requirements of Schr\"odinger invariance, the
implications of which we work out for general $n$-point functions.
The same integral for three-point function appeared in  \cite{Henkel:2003},
where these nonrelativistic three-point functions
where identified with response functions in Martin-Siggia-Rose theory,
but the integral has not been explicitly evaluated there.

The paper is organized as follows. In section 2 we
briefly
review the necessary results from non-relativistic 
conformal field theories and
discuss the constraints that Schr\"odinger symmetry places
on $n$-point functions of primary operators.
In section 3 we compute the bulk-to-bulk and bulk-to-boundary propagators
for a scalar field. In section 4 we compute three- and
four-point functions on the gravity side
and compare them with the corresponding correlators in NRCFT.
In appendix we explicitly evaluate the gravity bulk integrals which appear.

\bigskip

{\bf Note Added.}  While this paper was in preparation the
paper~\cite{Fuertes} appeared which also considers  three-point
function in non-relativistic AdS/CFT.

\section{Correlation Functions With Schr\"odinger Symmetry}

In this section we first review,
following~\cite{Son:2007}, a few of the most essential
features of the Schr\"odinger symmetry.
We then address the question of what can be said about
the structure of general $n$-point functions just based
on requiring invariance under this symmetry group,
generalizing some of the discussion of~\cite{Henkel:1993, Ginsparg}.

The generators of the Schr\"odinger group are
the number operator (or the `mass' operator) $M$,
the dilatation operator $D$ which generates
scale transformations
\begin{equation}
D : (\vec{x},t) \to (\lambda \vec{x}, \lambda^2 t)\,,
\end{equation}
the momentum and energy operators $P_i$ and $H$
which generate space and time translations
\begin{equation}
P_i:  \vec{x} \to \vec{x} + \vec{a}, \qquad H : T \to t + a\,,
\end{equation}
the angular momentum operators $M_{ij}$
which generate spatial rotations
\begin{equation}
M_{ij}:  \vec{x} \to R \vec{x}\,,
\end{equation}
the generators $K_i$ of Galilean boosts
\begin{equation}
K_i: \vec{x} \to \vec{x} - \vec{v} t\,,
\end{equation}
and finally the generator $C$ of special conformal transformations
\begin{equation}
C : (\vec{x},t) \to \left( \frac{\vec{x}}{1+at},\frac{t}{1 + a t}\right)\,,
\end{equation}
which can alternatively be expressed as $C = T_1 H T_1$ in terms of
\begin{equation}
T_1: (\vec{x},t) \to \left( \frac{\vec{x}}{t}, \frac{1}{t} \right)\,.
\end{equation}

For an operator $\mathcal{O}$
of definite conformal dimension $\Delta_{\mathcal{O}}$
we have~\cite{Son:2007}
\begin{equation}
\begin{aligned}
{}[D, [K_i, \mathcal{O}]]&=i(\Delta_\mathcal{O}-1)[K_i,
\mathcal{O}]\,,\\
[D, [C, \mathcal{O}]]&=i(\Delta_\mathcal{O}-2)[C, \mathcal{O}]\,.
\end{aligned}
\end{equation}
Since $K_i$ and $C$ evidently lower the dimension of any operator
we can define a primary operator to be one which satisfies the conditions
\begin{equation}\label{primary}
[K_i, \mathcal{O}]=0,\qquad [C, \mathcal{O}]=0\,.
\end{equation}
Starting with a primary operator one can build up a tower of
descendants by repeated commutation with $H$ and/or $P_i$, which
always raises the dimension.

Let us now consider what can be said in general about
the structure of an $n$-point correlation function
\begin{equation}
A_n(1,\ldots,n)
\equiv
\langle\mathcal{O}_1(\vec{x}_1,t_1)\cdots\mathcal{O}_n(\vec{x}_n,t_n)\rangle
\end{equation}
of primary operators.
First we can use translation invariance to reduce the $n$ coordinates
$(\vec{x}_i,t_i)$ to $(n-1)$ independent coordinates $(\vec{x}_{ij},
t_{ij}),$ where $t_{ij} = t_i-t_j$ and $\vec{x}_{ij}=(\vec{x}_i-\vec{x}_j)$.
Since the $(n-1)$ time variables are automatically invariant under
the $K_i$ we can build conformally invariant variables from them
just as in more familiar relativistic CFTs.  Specifically, scale invariance
implies that only ratios such as $t_{ij}/t_{kl}$ may appear, while
invariance under special conformal invariance allows only
the familiar cross-ratios of the form $t_{ij} t_{kl}/t_{ik} t_{jl}$,
of which $n(n-3)/2$ are independent \cite{Ginsparg}.

Now consider the $(n-1)$ vectors $\vec{x}_{ij}$, from which we can
build a total of $n(n-1)/2$ independent scalars $\vec{x}_{ij} \cdot
\vec{x}_{kl}$. Then we use $(n-1)$ Galilean boosts
similar to the procedure in
\cite{Ginsparg}
to reduce this number to $(n-1)(n-2)/2$. Actually the resulting
independent conformally invariant variables can be parametrized as
\begin{equation}
v_{ij} =
\frac{(\vec{x}_{in} t_{jn} -
\vec{x}_{jn} t_{in})^2}{2t_{ij} t_{in} t_{jn}}
={1 \over 2}
\left(\frac{x^2_{jn}}{t_{jn}}-\frac{x^2_{in}}{t_{in}}+\frac{x^2_{ij}}{t_{ij}} 
\right), ~~~~~
i<j<n\,,
\label{vexpr}
\end{equation}
and this is the form in which they will appear naturally from the AdS
calculations in the next section.
One can easily check the $v_{ij}$ are indeed
invariant under all of the Schr\"odinger group generators.

So in general an $n$-point function will always be allowed to
have arbitrary functional dependence on
a total of $n^2-3n+1$ Shr\"odinger-invariant variables.
The functional dependence on the remaining, non-conformally
invariant variables, can be determined by solving the
analogue of the conformal Ward identities, which are differential
equations expressing the constraints of the symmetry on
correlation functions.

For example it is well-known~\cite{Henkel:1993} that the 2-point
function is completely fixed up to an overall constant to the form
\begin{equation}
A_2(1,2) =c\,\delta_{\Delta_1,\Delta_2}
t^{-\Delta_1}_{12}e^{\frac{i M}{2}\frac{x^2_{12}}{t_{12}}}\,.
\label{2-point}
\end{equation}
The non-relatvistic 3-point function has been
shown~\cite{Henkel:1993} to be determined as
\begin{equation}
A_3(1,2,3)=
\prod_{i<j}t^{\Delta/2-(\Delta_i+\Delta_j)}_{ij}
e^{i(\frac{M_1}{2}\frac{x^2_{13}}{t_{13}}+
\frac{M_2}{2}\frac{x^2_{23}}{t_{23}})}F(v_{12})
\label{3-point}
\end{equation}
where $F$ is an arbitrary function.
Analagously, we find that the general form of the 4-point function is
\begin{equation}
\label{4-point}
A_4(1,2,3,4)=\prod_{i<j}t^{\Delta/6-(\Delta_i+\Delta_j)/2}_{ij}
e^{i(\frac{M_1}{2}\frac{x^2_{14}}{t_{14}}+
\frac{M_2}{2}\frac{x^2_{24}}{t_{24}}+\frac{M_3}{2}\frac{x^2_{34}}{t_{34}})}
F(\frac{t_{12}t_{34}}{t_{14}t_{23}},\frac{t_{12}t_{34}}{t_{13}t_{24}},
v_{12},v_{13},v_{23})\,
\end{equation}
where $\Delta=\sum_i \Delta_i$.
It is a simple exercise to check that this satisfies the
relevant conformal Ward identities.

\section{Non-relativistic AdS/CFT}

The metric with Schr\"odinger group isometry
constructed
in~\cite{D.T.Son:2008,McGreevy:2008} is
\begin{eqnarray}\label{metric}
\begin{aligned}
ds^2=L^2(-\frac{dt^2}{r^4}+\frac{2d\xi
dt+d\vec{x}^2}{r^2}+\frac{dr^2}{r^2})\,,
\end{aligned}
\end{eqnarray}
where $\vec{x}=x_i$ and $i=1,2 \ldots d.$
Let us consider a massive scalar field in this background.
The wave equation is
\begin{equation}\label{EOM}
(\nabla^2- m_0^2) \phi = \left(r^{d+3}\partial_r ( \frac{1}{
r^{d+1}}\partial_r)+r^2(2\partial_{\xi}\partial_t +
r^{-2}\partial^2_{\xi}+\partial^2_{i})-m_0^2
\right)\phi(r,\xi,x_i,t)=0\,.
\end{equation}
Because $\xi$ is a compact direction we can use

\begin{equation}
\label{xi}
\phi(r,\xi,x_i,t) = e^{  i M
\xi } \phi_{M}(r, x_i, t)\,
\end{equation}
to rewrite the equation~(\ref{EOM})
on the Fourier modes as
\begin{equation}\label{EOM2}
\begin{aligned}
&\left(r^{d+3}\partial_r ( \frac{1}{ r^{d+1}})+r^2(2iM\partial_t
+\partial^2_{i})-m^2 \right)\phi_{M}(r, x_i, t)=0\,,
\end{aligned}
\end{equation}
where $m^2=M^2+m^2_0$.
In order to make the equation more symmetric
we can introduce a new coordinate $\eta$ by
\begin{equation}
\phi_{M}(r, x_i, t)=\int d\eta\, e^{-iM\eta} \psi(r,\eta,x_i,t)\,.
\label{ft}
\end{equation}
This is the same trick that was introduced
in~\cite{Henkel:2003},
where the free massive Schr\"odinger equation
was reduced to a non-massive Klein-Gordon equation
via Fourier transform with respect to the mass.

If $\psi(r,\eta=\pm \infty,x_i,t) \to 0$ we can integrate by parts
so that~(\ref{EOM2}) becomes effectively
\begin{eqnarray}\label{EOM3}
\left(r^{d+3}\partial_r ( \frac{1}{
r^{d+1}}\partial_r)+r^2(2\partial_{\eta}\partial_t
+\partial^2_{i})-m^2 \right)\psi(r,\eta,x_i,t)=0\,,
\end{eqnarray}
and we can further simplify the equation by introducing
the $\chi$ coordinates
according to
\begin{equation}
\label{chi}
\begin{aligned}
t&=\sqrt{\frac{1}{2}}(\chi_0 - i \chi_{n+1})\,,\\
\eta&=\sqrt{\frac{1}{2}}(\chi_0 +  i\chi_{n+1})\,,\\
 x_i&=\chi_i\,.
\end{aligned}
\end{equation}

Then the wave equation~(\ref{metric}) becomes identical to
that of a scalar field in a Euclidean AdS background
\begin{eqnarray}\label{EOM4}
\left(r^{d+3}\partial_r ( \frac{1}{
r^{d+1}}\partial_r)+
r^2 (\partial^2_{\chi_i}+
\partial^2_{\chi_0}+
\partial^2_{\chi_{n+1}}
)-m^2
\right)\psi(r,\chi_i)=0\,.
\end{eqnarray}

We will use exactly same strategy in the next section to
compute the bulk-to-boundary and bulk-to-bulk propagators in the
background~(\ref{metric}).

\subsection{The bulk-to-boundary propagator and two-point function}

The boundary of the background (\ref{metric}) is at $r=0$ and
the generator associated with translations along the compact
$\xi$ direction idenfied as the mass operator $M= i \partial_\xi$,
So a $(d+3)$-dimensional
bulk theory is dual to a $(d+1)$-dimension boundary
theory~\cite{McGreevy:2008,D.T.Son:2008}.

Near the boundary, solutions of the scalar wave equation behave like
\begin{equation}
\phi \to r^{(d+2)-\Delta} \phi_0
\end{equation}
where
\begin{eqnarray}\label{Delta}
\Delta=1+\frac{d}{2} + \sqrt{\left(1+\frac{d}{2} \right)^2 + m^2_0 + M^2}
\end{eqnarray}
is related to the scaling dimension of the source
$\mathcal{O}$ in the boundary theory.
We will use the usual AdS/CFT recipe~\cite{witten:1998,gkp}
to calculate correlation functions of $\mathcal{O}$.

In order to compute the two-point function of the boundary operators
we have to calculate the on-shell action of a massive
scalar field for a solution of the classical
equation of motion
\begin{equation}
\nabla^2 \phi_M=m_0^2 \phi_M
\label{waveeq}
\end{equation}
subject to the boundary condition
\begin{equation}
\label{waeq} \lim_{r \to 0} \phi_M(r, t, \vec{x})=
r^{(d+2)-\Delta} \phi_0 (t, \vec{x})\,,
\end{equation}
where $\Delta$ is given in~(\ref{Delta}).
We used the compactness of $\xi$ (\ref{xi})
in order not to write  the explicit $\xi$ dependence
focusing on a Fourier mode $\phi_M$.

The relevant solution of~(\ref{waveeq}) is given by
\begin{eqnarray}\label{boundary propagator}
\phi_M(r,\vec{x},t)=\int d\vec{x}_1dt_1
K(r,\vec{x},t; \vec{x}_1, t_1)\phi_0(\vec{x}_1,t_1)\,,
\end{eqnarray}
where $K(r,\vec{x},t; \vec{x}_1, t_1)$
is the bulk-to-boundary propagator for metric (\ref{metric})
\begin{eqnarray}\label{propagator2}
\begin{aligned}
K(r, \vec{x},t; \vec{x}_1, t_1)&=\frac{i( \frac{M}{2}
)^{\Delta-1}e^{-i\pi
\Delta/2}}{\pi^{\frac{d}{2}}\Gamma(\Delta-(\frac{d}{2}+1))}
  \theta(t-t_1)
\left(\frac{r}{t-t_1}\right)^{\Delta}
e^{{i\over 2}M(1+i\epsilon){\frac{r^2+(x-x_1)^2}{(t-t_1)}}}\,,
\end{aligned}
\end{eqnarray}
where $\epsilon$ is the regulator.
As expected this heat kernel is the solution
of~(\ref{waveeq}) which in the limit $r \to 0$ behaves as
a delta function
\begin{eqnarray}
\begin{aligned}
r^{\Delta-(d+2)}K(r,\vec{x},t; \vec{x}_1, t_1)\to
\delta^d(x-x_1)\delta(t-t_1)\,.
\end{aligned}
\end{eqnarray}

Let us now show how the bulk-to-boundary propagator~(\ref{propagator2})
may be derived by using a trick similar to the one described in the
previous section.
Using~(\ref{xi}) and~(\ref{ft}), we get the following representation for
the bulk-to-boundary propagator,
\begin{equation}\label{boundary propagator1}
K(r,\vec{x},t; \vec{x}_1, t_1)=\int d\eta\,
e^{-iM\eta} K(r,\eta,\vec{x},t;\vec{x}_1, t_1),
\end{equation}
where $K(r,\eta,\vec{x},t;\vec{x}_1, t_1)$ is the bulk-to-boundary
propagator in Euclidean AdS space~(\ref{EOM4})~\cite{witten:1998,D.Freedman:1998,freedmanreview}
\begin{equation}
K(r,\eta,\vec{x},t;\vec{x}_1, t_1)
=c_{\Delta}
\left(\frac{r}{r^2+2(t-t_1)\eta+(x-x_1)^2}\right)^{\Delta}\,,
\end{equation}
with $\Delta$ given by~(\ref{Delta}) and $c_{\Delta}=\frac{i \Gamma(\Delta)}{\pi^{1+\frac{d}{2}}\Gamma(\Delta-(\frac{d}{2}+1))}$.

The
integral~(\ref{boundary propagator1}) is not well-defined as
it stands, so we
introduce the usual regulator $\epsilon$ by $t\rightarrow
t(1-i\epsilon)$
which is equivalent to taking
$t \to - i t$ as was done in~\cite{Henkel:2003}.  This leads to
integral representation of $K(r, \vec{x},t;
\vec{x}_1, t_1)$
\begin{equation}
c_{\Delta}
\int_{\bf R} d\eta~e^{-iM\eta}
\left(\frac{r}{r^2+2  (t-t_1)(1-i\epsilon)\eta+(x-x_1)^2}\right)^{\Delta}\,.
\end{equation}
It will become useful for computing higher point
correlation functions in the next section.
The result~(\ref{propagator2}) follows after performing the
$\eta$ integral here. 

Returning now to the solution~(\ref{boundary propagator}), we find
that when evaluating the action of this solution only a boundary
term contributes
\begin{equation}
S[\phi] =\lim_{r \to 0} \int d^{d} \vec{x} dt~ r^{-1-d} ~\phi
\partial_r \phi\,,
\end{equation}
and in this manner we find
\begin{equation}
S[\phi] =a_{\Delta} \int d^{d} \vec{x_1} dt_1  d^d \vec{x_2} dt_2
~\phi_0(\vec{x_1},t_1)\theta(t_1-t_2) \left( \frac{1}{t_1-t_2}
\right)^\Delta e^{i\frac{M}{2}\frac{(x_1-x_2)^2}{(t_1-t_2)}}
\phi_0(\vec{x_2},t_2),
\end{equation}
where\footnote{The careful
treatment of the
$r\rightarrow 0$ limit amounts to multiplying this result by
an additional factor of
$\frac{2\Delta-(d+2)}{\Delta}$
~\cite{D.Freedman:1998}. }
\begin{equation}
a_{\Delta}=\frac{i \Delta (\frac{M}{2})^{\Delta-1}
e^{-i \pi \Delta/2}
}{\pi^{\frac{d}{2}}
\Gamma(\Delta-(\frac{d}{2}+1))}\,.
\end{equation}
Thus the boundary action correctly reproduces the two-point function
in non-relativistic CFT (\ref{2-point})
\begin{equation}
A_2(1,2)
=\frac{i \Delta (\frac{M}{2})^{\Delta-1}e^{-i \pi \Delta/2} }
{\pi^{\frac{d}{2}}\Gamma(\Delta-(\frac{d}{2}+1))}\theta(t_1-t_2)
 \left( \frac{1}{t_1-t_2} \right)^\Delta
e^{i\frac{M}{2}\frac{(x_1-x_2)^2}{(t_1-t_2)}}.
\end{equation}
This was also found in~\cite{McGreevy:2008}.

\subsection{The bulk-to-bulk propagator}

The bulk-to-bulk propagator $G_M(r,\vec{x},t;r_1,\vec{x}_1,t_1)$
for a scalar field 
in background (\ref{metric})
satisfies
\begin{multline}
\left(r^{d+3}\partial_r ( \frac{1}{
r^{d+1}}\partial_r)+r^2(2iM\partial_t
+\partial^2_{i})-(m_0^2+M^2)\right)
G_M(r,\vec{x},t;r_1,\vec{x}_1,t_1)\\
=r^{d+3}\delta(r-r_1)\delta^d(\vec{x}-\vec{x}_1)\delta(t-t_1)\,.
\end{multline}

Here we again used the compactness of $\xi$ and wrote
an equation for the corresponding Fourier mode, in terms of which the full
bulk-to-bulk propagator is
\begin{equation}
G(r,\xi,\vec{x},t;r_1,\xi_1,\vec{x}_1, t_1)=
e^{i M (\xi-\xi_1)} G_M(r,\vec{x},t;r_1,\vec{x}_1, t_1).
\end{equation}

If we  introduce the coordinate $\eta$ by\footnote{It might appear that we
have introduced one more unwanted parameter $M_1$, but because
$G(r,\eta,\vec{x},t;r_1,\eta_1,\vec{x}_1, t_1)$ is
invariant under
translations in
$\eta$, the integral will force $M=M_1$. That is why
there is a $\delta(M-M_1)$ on the left hand side.}
\begin{equation}
\label{bulkprop}
\delta(M-M_1)G_M(r,\vec{x},t;r_1, \vec{x}_1, t_1)=\int d\eta d\eta_1
\,e^{-iM\eta+iM_1\eta_1} G_M(r,\eta,\vec{x},t;r_1,\eta_1,\vec{x}_1, t_1)\,,
\end{equation}
we then find that $G_M(r,\eta,\vec{x},t;r_1,\eta_1,\vec{x}_1, t_1)$
satisfies
\begin{multline}\label{bulk-bulk2}
\left(r^{d+3}\partial_r ( \frac{1}{
r^{d+1}}\partial_r)+r^2(2\partial_{\eta}\partial_t
+\partial^2_{i})-m^2\right)G_M(r,\eta,\vec{x},t;r_1,\eta_1,\vec{x}_1, t_1)\\
=r^{d+3}
\delta(\eta-\eta_1)\delta(r-r_1)\delta^d(\vec{x}-\vec{x}_1)\delta(t-t_1)\,.
\end{multline}

This in turn is nothing other than
the equation for bulk-to-bulk propagator in $AdS$,
whose solution
is~\cite{D'Hoker:1998,freedmanreview}
\begin{equation}
G_M(r,\eta,\vec{x},t;r_1,\eta_1,\vec{x_1},t_1)=\tilde{C}_{\Delta}
(2z)^{\Delta}{}_2F_1 \left(\frac{\Delta}{2},
\frac{\Delta}{2}+\frac{1}{2};\Delta-{d \over 2},z^2 \right)\,,
\end{equation}
where
\begin{align}
\tilde{C}_{\Delta}&=
\frac{i\Gamma(\Delta)\Gamma(\Delta-\frac{d}{2}-\frac{3}{2})}{(4\pi)^{(d+3)/2}
\Gamma(2\Delta-d-1)}\,,\\
z&=
\frac{2rr_1}{r^2+r_1^2+(\vec{x}-\vec{x_1})^2+2 (\eta-\eta_1)(t-t_1)}\,.
\end{align}

By changing variables we find that bulk-to-bulk propagator
may be expressed as
\begin{equation}
\begin{aligned}
\label{bulk-bulk}
&G_M(r,\vec{x},t;r_1,
\vec{x}_1, t_1)=\\
&\tilde{C}_{\Delta}
e^{{i \over 2}M(1+i\epsilon)\frac{y^2}{t-t_1}}
\left(\frac{2rr_1}{t-t_1}\right)^{\Delta}
\int_{\mathcal{C}} du\, e^{-iMu} u^{-\Delta}{}_2F_1 \left(\frac{\Delta}{2},
\frac{\Delta}{2}+\frac{1}{2};\Delta-{d \over 2};
\left(\frac{rr_1}{(t-t_1)u}\right)^2
\right)
\end{aligned}
\end{equation}
where $y^2=r^2+r^2_1+(\vec{x}-\vec{x}_1)^2$ and 
$\mathcal{C}$ is ${\bf R}+ {y^2 \over 2(t-t_1)}i \epsilon$.

Using
\begin{equation}
{}_2F_1(a, b;c;z)=\sum^\infty_{n=0}\frac{(a)_{n}(b)_n}{(c)_{n}n!}z^n
\end{equation}
we have
\begin{equation}
\begin{aligned}
&\int_{\mathcal{C}} du\, e^{-iMu} u^{-\Delta}{}_2F_1(a,
b;c;(k/u)^2)=\\
&= \sum^\infty_{n=0}\frac{(a)_{n}(b)_n}{(c)_{n}n!}(k^2)^n
\int_{\mathcal{C}} du~ e^{-iMu} u^{-(\Delta+2n)}=\\
&=2 \pi M^{\Delta-1} e^{-i \pi \Delta/2}
\sum^\infty_{n=0}\frac{(a)_{n}(b)_n }{(c)_{n}n!} { (-iM k)^{2n} \over
\Gamma(\Delta+2n)}=\\
&={2 \pi M^{\Delta-1} e^{-i \pi \Delta/2} \over \Gamma({\Delta})}
{}_2F_3(a,b;c,{1 \over 2}+{\Delta \over 2},{\Delta \over 2};
-{M^2 k^2 \over 4})
\,.
\end{aligned}
\end{equation}

Putting everything together we find that the bulk-to-bulk propagator
is given by
\begin{equation}
\label{bulkintegralresult}
G_M(r,\vec{x},t;r_1, \vec{x}_1, t_1)=
K_{\Delta} ~
\theta(t-t_1)
e^{\frac{i M(1+i \epsilon)}{2}\frac{y^2}{t-t_1}}
\left({r r_1 \over t-t_1  }\right)^{{d \over 2}+1}
J_{\Delta-{d \over 2}-1} \left({ M r r_1 \over t-t_1}\right),
\end{equation}
where $J$ is the Bessel function
and the constant is
 \begin{equation}
K_{\Delta}=\frac{\pi \tilde{C}_{\Delta} M^{d/2} 2^{2 \Delta -d/2}
e^{-i\pi\Delta/2}
\Gamma(\Delta-d/2)}
{\Gamma(\Delta)}.
\end{equation}

\section{Higher-Point Correlation Functions}

\subsection{Three-point functions}

Now let us calculate boundary
three-point functions. We consider bulk interaction vertices of the form
$L_1=\phi^3$ and $L_2= \phi \partial_\mu \phi \partial^\mu \phi$.
The corresponding three-point functions are
respectively
\begin{align}
A_3(1,2,3)&=\int \frac{dr d^dxdt d\xi}{r^{d+3}}
{K}^*_1(r,\xi,\vec{x},t; \vec{x}_1,t_1)
{K}^*_2(r,\xi,\vec{x},t; \vec{x}_2,t_2)\
{K}_3(r,\xi,\vec{x},t; \vec{x}_3, t_3)\,,\\
\label{3-point2}
A_3'(1,2,3)&=\int
\partial_{\mu}{K}^*_1\partial^{\mu}{K}^*_2{K}_3\,.
\end{align}

Using the representation~(\ref{boundary propagator1})
of the bulk-to-boundary
propagator,
changing variables and using number conservation
(which comes from the $\xi$ integral)
we find
\begin{multline}
A_1(1,2,3) =c_1\int d\alpha' d\beta' e^{-iM_1\alpha'}
e^{-iM_2\beta'}\int \frac{drd^dxdtd\gamma}{r^{d+3}}
\left(\frac{r}{r^2+2(t-t_1)(\gamma-\alpha')+(x-x_1)^2}\right)^{\Delta_1}
\\
\times \left(\frac{r}{r^2+2(t-t_2)(\gamma-\beta')+(x-x_2)^2}\right)^{\Delta_2}
\left(\frac{r}{r^2+2(t-t_3)\gamma+(x-x_3)^2}\right)^{\Delta_3},
\end{multline}
where $c_1=\bar{c}_{\Delta_1} \bar{c}_{\Delta_2} {c}_{\Delta_3}.$

Using the known relativistic AdS/CFT~\cite{D'Hoker:1998} result and introducing the regulator,
we can perform the integral to arrive at
\begin{equation}\label{3-point1}
A_1(1,2,3)=a_1(t_{12})^{-\Delta_{12,3}/2}(t_{23})^{-\Delta_{23,1}/2}(t_{13})^{-\Delta_{13,2}/2}
e^{i(\frac{M_1}{2}\frac{x^2_{13}}{t_{13}}+\frac{M_2}{2}\frac{x^2_{23}}{t_{23}})}I
\end{equation}
where
\begin{equation}
a_1=-\frac{\Gamma[\frac{1}{2}(\Delta-(d+2))]
M^{\Delta_1-1}_1M^{\frac{\Delta_{23,1}}{2}-1}_2 e^{-i\pi \Delta/4}}
{2^{\Delta/2-1}\pi^{d}\Gamma[\Delta_1-\frac{d}{2}-1]\Gamma[\Delta_2-\frac{d}{2}-1]\Gamma[\Delta_3-\frac{d}{2}-1]}
\end{equation}
and 
\begin{align}\label{integral}
I&=\frac{1}{C}\int_{\mathcal{C}_1} du\int_{\mathcal{C}_2} dw
e^{-iM_1u} e^{-iM_2w}\frac{1}{(u-w+v_{12}(1+i\epsilon))^{\Delta_{12,3}/2}}
\frac{1}{w^{\Delta_{23,1}/2}u^{\Delta_{13,2}/2}}\\
&=\theta(t_{23})\theta(t_{13})
B(\frac{\Delta_{12,3}}{2},\frac{\Delta_{13,2}}{2})
\Phi_1(\frac{\Delta_{12,3}}{2},\frac{\Delta_{23,1}}{2}-1,\Delta_1,-\frac{M_1}{M_2},iM_1v_{12})\,
\end{align}
where the contour $\mathcal{C}_j$ is $({\bf
R}+i\frac{\vec{x}_{j3}^2}{2t_{j3}}\epsilon)$ and
\begin{align}
v_{12}&=\frac{1}{2}(\frac{x^2_{12}}{t_{12}}+\frac{x^2_{23}}{t_{23}}-\frac{x^2_{13}}{t_{13}})\,,\\
\Delta&=\sum_i \Delta_i, \qquad \Delta_{ij,k}=\Delta_i+\Delta_j-\Delta_k\,,\\
C&=\frac{4\pi^2M^{\Delta_1-1}_1M^{\frac{\Delta_{23,1}}{2}-1}_2e^{-\frac{i\pi}{4}\Delta}}{\Gamma[\frac{\Delta_{12,3}}{2}]\Gamma[\frac{\Delta_{23,1}}{2}]
\Gamma[\frac{\Delta_{13,2}}{2}]}\,.
\end{align}
The beta function $B(\nu,\lambda)$ and the confluent hypergeometric
$\Phi_1(\alpha,\beta,\gamma,x,y)$ are defined in~(\ref{betafunction})
and~(\ref{hypergeometric}) respectively.

Similarly, for the second type of interaction we find
\begin{equation}
A_2(1,2,3)=(a_2-a_1M_1M_2)(t_{12})^{-\Delta_{12,3}/2}(t_{23})^{-\Delta_{23,1}/2}(t_{13})^{-\Delta_{13,2}/2}
e^{i(\frac{M_1}{2}\frac{x^2_{13}}{t_{13}}+\frac{M_2}{2}\frac{x^2_{23}}{t_{23}})}I
\end{equation}
where
\begin{equation}
a_2=a_1\big[\Delta_2\Delta_3+\frac{1}{2}(d+2-\Delta)(\Delta_{23,1})\big]
\end{equation}
and $I$ is the same integral as~(\ref{integral}).
The details of the evaluation of~(\ref{integral}) are shown in an appendix. We note that this is the same integral which 
appeared in \cite{Henkel:2003} where the nonrelativistic three-point functions
in CFT where identified with response functions in Martin-Siggia-Rose theory,
but the integral has not been explicitly evaluated there.

\subsection{Four-point functions}

Finally, we consider the four-point function arising from a
bulk quartic interaction $\phi^4$
\begin{equation}
A_1(1,2,3,4)=\int 
{K}^*_1(r,\vec{x},t; \vec{x}_1,
t_1){K}^*_2(r,\vec{x},t; \vec{x}_2,
t_2){K}^*_3(r,\vec{x},t; \vec{x}_3,
t_3){K}_4(r,\vec{x},t; \vec{x}_4, t_4)\,.
\end{equation}
Performing the same tricks as above we find
\begin{equation}
A_1(1,2,3,4) =\int d\alpha_1 d\alpha_2 d\alpha_3\,
e^{-iM_1\alpha_1}e^{-iM_2\alpha_2}e^{-iM_3\alpha_3} I_1,
\end{equation}
where
\begin{multline}
I_1=b_1 \int \frac{drd^dxdtd\mu}{r^{d+3}}
\left(\frac{r}{r^2+(x-x_1)^2+2(t-t_1)(\mu-\alpha_1)}\right)^{\Delta_1}\\
\times
\left(\frac{r}{r^2+(x-x_2)^2+2(t-t_2)(\mu-\alpha_2)}\right)^{\Delta_2}\\
\times
\left(\frac{r}{r^2+(x-x_3)^2+2(t-t_3)(\mu-\alpha_3)}\right)^{\Delta_3}
\left(\frac{r}{r^2+(x-x_4)^2+2(t-t_4)\mu}\right)^{\Delta_4}
\end{multline}
is the same integral as in the relativistic
AdS/CFT case~\cite{W.Mueck:1998,freedmanreview}
and $b_1=\bar{c}_{\Delta_1} \bar{c}_{\Delta_2}  \bar{c}_{\Delta_3} c_{\Delta_4}$.

Let us consider as an example the simple case,
$\Delta_1=\Delta_2=\Delta_3=\Delta_4=\Delta$, for which it
is known~\cite{W.Mueck:1998} that
\begin{equation}
I_1=\frac{\Gamma(2\Delta-\frac{d}{2}-1)}
{\Gamma(2\Delta)}\frac{2\pi^{(d+2)/2}}{(\chi^2_{12}\chi^2_{34})^{\Delta}}
\int^{\infty}_0 dz\, {}_2F_1(\Delta, \Delta;2\Delta;
1-\frac{(\eta+\zeta)^2}{(\eta \zeta)^2}-\frac{4}{\eta
\zeta}\sinh^2z)
\end{equation}
where
\begin{equation}
\eta=\frac{\chi_{12}\chi_{34}}{\chi_{14}\chi_{23}},
\qquad \zeta=\frac{\chi_{12}\chi_{34}}{\chi_{13}\chi_{24}}
\end{equation}
are conformal cross-ratios
and
\begin{equation}
\chi^2_{ij}=(x_{ij}^2+2 t_{ij}\alpha_{ij}).
\end{equation}

Again introducing the regulator and making the change of variables
 $u_i=\alpha_i+\frac{x^2_{i4}}{2t_{i4}}(1+i\epsilon)$, we have
\begin{equation}
\chi^2_{i4}=2 t_{i4}(1-i\epsilon)u_i\,,
\qquad \chi^2_{ij}=2 t_{ij}(1-i\epsilon)(u_{ij}+v_{ij}(1+i\epsilon))  \qquad \text{for $i,j<4$}\,,
\end{equation}
where $v_{ij}$ is defined in (\ref{vexpr}) and $\eta$ and $\zeta$ change similarly.

Finally then we arrive at
\begin{equation}
A_1(1,2,3,4)
=\frac{\Gamma(2\Delta-\frac{d}{2}-1)}{\Gamma(2\Delta)}\frac{2\pi^{(d+2)/2}}{(4t_{12}t_{34})^{\Delta}}
e^{i(\frac{M_1}{2}\frac{x^2_{14}}{t_{14}}+\frac{M_2}{2}\frac{x^2_{24}}{t_{24}}+\frac{M_3}{2}\frac{x^2_{34}}{t_{34}})}I_2
\end{equation}
in terms of
\begin{multline}
I_2=\int_{\mathcal{C}_1} du_1 \int_{\mathcal{C}_2} du_2
\int_{\mathcal{C}_3} du_3
e^{-iM_1u_1}e^{-iM_2u_2}e^{-iM_3u_3}\\
({u_3(u_{12}+v_{12}(1+i\epsilon))})^{-\Delta} \int^{\infty}_0 dz
{}_2F_1(\Delta, \Delta;2\Delta, 1-\frac{(\eta+\zeta)^2}{(\eta
\zeta)^2}-\frac{4}{\eta \zeta}\sinh^2z)\,,
\end{multline}
where the contour $\mathcal{C}_j$ is $({\bf
R}+i\frac{x^2_{j4}}{2t_{j4}}\epsilon)$.
We see the result is consistent with
the general four-point function~(\ref{4-point}).

We could treat the scalar exchange diagram similarly by
first performing the Fourier transform and then doing the integral over
the AdS bulk, leading to
\begin{equation}
A_2(1,2,3,4) =\int d\alpha' d\beta' d\gamma' d\mu'
e^{-iM_1\alpha'}e^{-iM_2\beta'}e^{iM_3\gamma'}e^{iM_4\mu'} I_4\,,
\end{equation}
where the integral $I_4$ is well-known from the original
AdS/CFT literature~\cite{Hong Liu:1998}
\begin{multline}
I_4=\int \frac{dzd^dxdtd\eta}{z^{d+3}}\frac{dwd^dyd\tau
d\eta'}{w^{d+3}}{K}_1(z,\eta,\vec{x},t;\alpha',\vec{x}_1,t_1){K}_3(z,\eta,\vec{x},t;\beta',\vec{x}_3,t_3)\\
G(z,\eta,\vec{x},t;w,\eta',\vec{y},\tau){K}_2(w,\eta',\vec{y},\tau;\gamma',\vec{x}_2,t_2)
{K}_4(w,\eta',\vec{y},\tau;\mu',\vec{x}_4,t_4)\,.
\end{multline}
After a change of variables, the integral could be brought to the
form (\ref{4-point}) as required by non-relativistic conformal invariance.

\section*{Acknowledgments}

We are grateful to  K.~Balasubramanian, 
A.~Jevicki, K.~Jin, C.~Kalousios, S.~Roy
and especially M.~Spradlin for useful conversations.
This work was supported in part by the US
Department of Energy under contract DE-FG02-91ER40688 
 and the US National Science Foundation under grant 
PHY-0643150 CAREER and PECASE.

\appendix

\section{Three-point function integral}
\label{the integral}

Here we consider the integral~(\ref{integral})
\begin{equation}
\int_{\mathcal{C}_1} du\int_{\mathcal{C}_2} dw~ e^{-iM_1u}
e^{-iM_2w}\frac{1}{(u-w+v_{12}(1+i\epsilon))^{\Delta_{12,3}/2}}
\frac{1}{w^{\Delta_{23,1}/2}u^{\Delta_{13,2}/2}}
\end{equation}
where the contour $\mathcal{C}_j$ is
$({\bf R}+i\frac{\vec{x}_{j3}^2}{2t_{j3}}\epsilon)$ for $\epsilon>0$.

Let us consider the case $t_{23}>0$ and $t_{12}>0$.
Combining the two denominators involving $u$ with
a Feynman parameter $z$ leads to
\begin{equation}
I = { \Gamma( {\Delta_{12,3} \over 2} + {\Delta_{13,2} \over 2} )
\over \Gamma(  {\Delta_{12,3} \over 2} ) \Gamma( {\Delta_{13,2}
\over 2} )} \int_0^1 dz\ z^{{\Delta_{12,3} \over 2} - 1} (1 - z)^{
{\Delta_{13,2} \over 2} -1} \int_{\mathcal{C}_2} dw \ e^{-i
M_2 w} w^{-\Delta_{23,1}/2} \ A\,,
\end{equation}
where
\begin{eqnarray}
\begin{aligned}
 A &= \int_{ \mathcal{C}_1} du  \ { e^{-i M_1 u} \over [u +
z (-w + v_{12}(1+i\epsilon))]^{ {\Delta_{12,3} \over 2} + {\Delta_{13,2} \over
2} }} \cr &= 2 \pi { (M_1)^{ {\Delta_{12,3} \over 2} +
{\Delta_{13,2} \over 2} - 1} \over \Gamma({\Delta_{12,3} \over 2} +
{\Delta_{13,2} \over 2} )} \exp \left[ i M_1 z (-w + v_{12}(1+i\epsilon))\right]
\exp\left[- {i \pi \over 2} \left( {\Delta_{12,3} \over 2} +
{\Delta_{13,2} \over 2}\right)\right]\,,
\end{aligned}
\end{eqnarray}
which we have evaluated by using the identity
\begin{equation}
\int_{ {\bf R} + i \alpha} du \frac{e^{-i M u} }{ (u +
z)^f} = 2 \pi \frac{M^{f-1} }{ \Gamma(f)} e^{i z M} e^{-i \pi f/2},
\qquad f > 0, M > 0, {\rm Im}(z) \geq 0\,.
\end{equation}
Noting that ${\Delta_{12,3} \over 2} + {\Delta_{13,2} \over 2} =
\Delta_1$, the original integral then becomes
\begin{equation}
 I = {2 \pi  (M_1)^{\Delta_1-1} e^{-i \pi \Delta_1/2}
\over \Gamma(  {\Delta_{12,3} \over 2} ) \Gamma( {\Delta_{13,2}
\over 2} )} \int_0^1 dz\ z^{{\Delta_{12,3} \over 2} - 1} (1 - z)^{
{\Delta_{13,2} \over 2} -1} e^{i M_1 z v_{12} (1+i \epsilon)} B
\end{equation}
where \begin{eqnarray}
\begin{aligned} B = \int_{\mathcal{C}_2} dw \ e^{-i M_2 w}
w^{-\Delta_{23,1}/2} e^{- i M_1 z w} = 2 \pi { (M_2 + z M_1)^{
{\Delta_{23,1} \over 2} - 1} \over \Gamma( { \Delta_{23,1} \over 2}
)} \exp\left[ - {i \pi \over 2} {\Delta_{23,1} \over 2} \right].
\end{aligned}
\end{eqnarray}
Putting everything together then gives
\begin{multline}
 I = {4 \pi^2 M_1^{\Delta_1 - 1} M_2^{ {\Delta_{23,1} \over 2} - 1} \over \Gamma(
{\Delta_{12,3} \over 2} ) \Gamma( {\Delta_{13,2} \over 2} ) \Gamma(
{\Delta_{23,1} \over 2} ) } \exp \left[ - {i \pi \over 4}
\Delta\right] \\
\times \int_0^1 dz \ z^{{\Delta_{12,3}
\over 2} - 1} (1 - z)^{ {\Delta_{13,2} \over 2} - 1} (1+
\frac{M_1}{M_2}z)^{ {\Delta_{23,1} \over 2} - 1} \exp[iM_1 z v_{12}(1+i\epsilon)]\,,
\end{multline}
and the final
integral can be evaluated thanks
to the identity~\cite{I.S.Gradsbteyn}
\begin{equation}
\int^1_0 x^{\nu-1}(1-x)^{\lambda-1}(1+\beta x)^{-\rho}e^{-\mu
x}dx=B(\nu,\lambda)\Phi_1(\nu,\rho,\lambda+\nu,-\beta,-\mu)\,,
\end{equation}
which leads to the advertised result~(\ref{integral}) in terms of
the beta function
\begin{equation}
\label{betafunction}
B(x,y)=\int^1_0 t^{x-1}(1-t)^{y-1}dt
\end{equation}
and the confluent hypergeometric function
function
\begin{equation}
\label{hypergeometric}
\Phi_1(\alpha,\beta,\gamma,x,y)
=\sum^\infty_{m,n=0}\frac{(\alpha)_{m+n}(\beta)_m}{(\gamma)_{m+n}m!n!}x^my^n\,.
\end{equation}


\begin{thebibliography}{99}

\bibitem{adscft}
  J.~M.~Maldacena,
  ``The large N limit of superconformal field theories and supergravity,''
  Adv.\ Theor.\ Math.\ Phys.\  {\bf 2}, 231 (1998)
  [Int.\ J.\ Theor.\ Phys.\  {\bf 38}, 1113 (1999)]
  [arXiv:hep-th/9711200].

\bibitem{gkp}
  S.~S.~Gubser, I.~R.~Klebanov and A.~M.~Polyakov,
  ``Gauge theory correlators from non-critical string theory,''
  Phys.\ Lett.\  B {\bf 428}, 105 (1998)
  [arXiv:hep-th/9802109].

\bibitem{witten:1998}
  E.~Witten,
  ``Anti-de Sitter space and holography,''
  Adv.\ Theor.\ Math.\ Phys.\  {\bf 2}, 253 (1998)
  [arXiv:hep-th/9802150].



\bibitem{applications}
  C.~P.~Herzog, A.~Karch, P.~Kovtun, C.~Kozcaz and L.~G.~Yaffe,
  JHEP {\bf 0607}, 013 (2006)
  [arXiv:hep-th/0605158].
$\bullet$
  S.~S.~Gubser,
  Phys.\ Rev.\  D {\bf 74}, 126005 (2006)
  [arXiv:hep-th/0605182].
$\bullet$
  C.~P.~Herzog,
  JHEP {\bf 0609}, 032 (2006)
  [arXiv:hep-th/0605191].
$\bullet$
  J.~J.~Friess, S.~S.~Gubser and G.~Michalogiorgakis,
  JHEP {\bf 0609}, 072 (2006)
  [arXiv:hep-th/0605292].
$\bullet$
  H.~Liu, K.~Rajagopal and U.~A.~Wiedemann,
  JHEP {\bf 0703}, 066 (2007)
  [arXiv:hep-ph/0612168].
$\bullet$
  D.~T.~Son and A.~O.~Starinets,
  Ann.\ Rev.\ Nucl.\ Part.\ Sci.\  {\bf 57}, 95 (2007)
  [arXiv:0704.0240 [hep-th]].
$\bullet$
  C.~P.~Herzog, P.~Kovtun, S.~Sachdev and D.~T.~Son,
  Phys.\ Rev.\  D {\bf 75}, 085020 (2007)
  [arXiv:hep-th/0701036].
$\bullet$
  S.~A.~Hartnoll, P.~K.~Kovtun, M.~Muller and S.~Sachdev,
  Phys.\ Rev.\  B {\bf 76}, 144502 (2007)
  [arXiv:0706.3215 [cond-mat.str-el]].
$\bullet$
  S.~A.~Hartnoll, C.~P.~Herzog and G.~T.~Horowitz,
  Phys.\ Rev.\ Lett.\  {\bf 101}, 031601 (2008)
  [arXiv:0803.3295 [hep-th]].
$\bullet$
  S.~S.~Gubser,
  Gen.\ Rel.\ Grav.\  {\bf 39}, 1533 (2007)
  [Int.\ J.\ Mod.\ Phys.\  D {\bf 17}, 673 (2008)].
$\bullet$
  S.~S.~Gubser and A.~Karch,
  arXiv:0901.0935 [hep-th].

  \bibitem{experimetal}
   K.M.O'Hara et al.,
   arXiv:cond-mat/0212463
   $\bullet$
    C.A.Regal, M.Greiner, D.S.Jin
    arXiv:cond-mat/0401554
    $\bullet$
    M. Bartenstein et al.,
    arXiv:cond-mat/0401109
    $\bullet$
    M.W. Zwierlein et al.,
    arXiv:cond-mat/0403049
    $\bullet$
    J. Kinast et al.,
    arXiv:cond-mat/0403540
    $\bullet$
    T. Bourdel et al.,
    arXiv:cond-mat/0403091




\bibitem{D.T.Son:2008}
  D.~T.~Son,
  ``Toward an AdS/cold atoms correspondence: a geometric realization of the
  Schroedinger symmetry,''
  Phys.\ Rev.\  D {\bf 78}, 046003 (2008)
  [arXiv:0804.3972 [hep-th]].

\bibitem{McGreevy:2008}
  K.~Balasubramanian and J.~McGreevy,
  ``Gravity duals for non-relativistic CFTs,''
  Phys.\ Rev.\ Lett.\  {\bf 101}, 061601 (2008)
  [arXiv:0804.4053 [hep-th]].


\bibitem{Maldacena}
    J.~Maldacena, D.~Martelli, Y.~Tachikawa
    ``Comments on string theory backgrounds with non-relativistic conformal symmetry.,''
   arXiv:0807.1100 [hep-th]

\bibitem{Adams}
    A.Adams, K.Balasubramanian, J.McGreevy
    ``Hot Spacetimes for Cold Atoms,''
   arXiv:0807.1111 [hep-th]

\bibitem{Herzog}
    C.P.~Herzog, M.~Rangamani, S.F.~Ross
    ``Heating up Galilean holography.,''
   arXiv:0807.1099 [hep-th]


\bibitem{NRcft}


  M.~Alishahiha, R.~Fareghbal, A.~E.~Mosaffa and S.~Rouhani,
  arXiv:0902.3916 [hep-th].
$\bullet$
  K.~M.~Lee, S.~Lee and S.~Lee,
  arXiv:0902.3857 [hep-th].
$\bullet$
  A.~Galajinsky and I.~Masterov,
  arXiv:0902.2910 [hep-th].
$\bullet$
  Y.~Nakayama,
  arXiv:0902.2267 [hep-th].
$\bullet$
  Y.~Nakayama, M.~Sakaguchi and K.~Yoshida,
  arXiv:0902.2204 [hep-th].
$\bullet$
  A.~Bagchi and R.~Gopakumar,
  arXiv:0902.1385 [hep-th].
$\bullet$
  A.~Akhavan, M.~Alishahiha, A.~Davody and A.~Vahedi,
  arXiv:0902.0276 [hep-th].
$\bullet$
  P.~Horava,
  arXiv:0901.3775 [hep-th].
$\bullet$
  M.~Alishahiha and A.~Ghodsi,
  arXiv:0901.3431 [hep-th].
$\bullet$
  A.~Donos and J.~P.~Gauntlett,
  arXiv:0901.0818 [hep-th].
$\bullet$
  S.~S.~Pal,
  arXiv:0901.0599 [hep-th].
$\bullet$
  U.~H.~Danielsson and L.~Thorlacius,
  arXiv:0812.5088 [hep-th].
$\bullet$
  M.~Taylor,
  arXiv:0812.0530 [hep-th].
$\bullet$
  A.~Adams, A.~Maloney, A.~Sinha and S.~E.~Vazquez,
  arXiv:0812.0166 [hep-th].
$\bullet$
  A.~Akhavan, M.~Alishahiha, A.~Davody and A.~Vahedi,
  arXiv:0811.3067 [hep-th].
$\bullet$
  Y.~Nakayama, S.~Ryu, M.~Sakaguchi and K.~Yoshida,
  JHEP {\bf 0901}, 006 (2009)
  [arXiv:0811.2461 [hep-th]].
$\bullet$
  M.~Rangamani, S.~F.~Ross, D.~T.~Son and E.~G.~Thompson,
  JHEP {\bf 0901}, 075 (2009)
  [arXiv:0811.2049 [hep-th]].
$\bullet$
  L.~Mazzucato, Y.~Oz and S.~Theisen,
  arXiv:0810.3673 [hep-th].
$\bullet$
  M.~Schvellinger,
  JHEP {\bf 0812}, 004 (2008)
  [arXiv:0810.3011 [hep-th]].
$\bullet$
  S.~A.~Hartnoll and K.~Yoshida,
  JHEP {\bf 0812}, 071 (2008)
  [arXiv:0810.0298 [hep-th]].
$\bullet$
  F.~L.~Lin and S.~Y.~Wu,
  arXiv:0810.0227 [hep-th].
$\bullet$
  D.~Yamada,
  arXiv:0809.4928 [hep-th].
$\bullet$
  C.~Duval, M.~Hassaine and P.~A.~Horvathy,
  arXiv:0809.3128 [hep-th].
$\bullet$
  P.~Kovtun and D.~Nickel,
  Phys.\ Rev.\ Lett.\  {\bf 102}, 011602 (2009)
  [arXiv:0809.2020 [hep-th]].
$\bullet$
  S.~Pal,
  arXiv:0809.1756 [hep-th].
$\bullet$
  S.~Sekhar Pal,
  arXiv:0808.3232 [hep-th].
$\bullet$
  S.~S.~Pal,
  arXiv:0808.3042 [hep-th].
$\bullet$
  S.~Kachru, X.~Liu and M.~Mulligan,
  Phys.\ Rev.\  D {\bf 78}, 106005 (2008)
  [arXiv:0808.1725 [hep-th]].
$\bullet$
  A.~V.~Galajinsky,
  Phys.\ Rev.\  D {\bf 78}, 087701 (2008)
  [arXiv:0808.1553 [hep-th]].
$\bullet$
  D.~Minic and M.~Pleimling,
  arXiv:0807.3665 [cond-mat.stat-mech].
$\bullet$
  J.~W.~Chen and W.~Y.~Wen,
  arXiv:0808.0399 [hep-th].
$\bullet$
  Y.~Nakayama,
  JHEP {\bf 0810}, 083 (2008)
  [arXiv:0807.3344 [hep-th]].
$\bullet$
  W.~Y.~Wen,
  arXiv:0807.0633 [hep-th].
$\bullet$
  M.~Sakaguchi and K.~Yoshida,
  JHEP {\bf 0808}, 049 (2008)
  [arXiv:0806.3612 [hep-th]].
$\bullet$
  J.~L.~B.~Barbon and C.~A.~Fuertes,
  JHEP {\bf 0809}, 030 (2008)
  [arXiv:0806.3244 [hep-th]].
$\bullet$
  W.~D.~Goldberger,
  arXiv:0806.2867 [hep-th].
$\bullet$
  M.~Sakaguchi and K.~Yoshida,
  arXiv:0805.2661 [hep-th].

\bibitem{Henkel:2003}
   M.~Henkel, J.~Unterberger,
  ``Schrodinger invariance and space-time symmetries,''
  arXiv:hep-th/0302187
  
  
   
\bibitem{Fuertes}
  C.~A.~Fuertes and S.~Moroz,
  ``Correlation functions in the non-relativistic AdS/CFT correspondence,''
  arXiv:0903.1844 [hep-th].
  
  

\bibitem{Son:2007}
  Y.~Nishida and D.~T.~Son,
  ``Nonrelativistic conformal field theories,''
  Phys.\ Rev.\  D {\bf 76}, 086004 (2007)
  [arXiv:0706.3746 [hep-th]].

\bibitem{Henkel:1993}
  M.~Henkel,
  ``Schrodinger invariance in strongly anisotropic critical systems,''
  J.\ Statist.\ Phys.\  {\bf 75}, 1023 (1994)
  [arXiv:hep-th/9310081].
  
\bibitem{Ginsparg}
  P.~H.~Ginsparg,
  ``APPLIED CONFORMAL FIELD THEORY,''
  arXiv:hep-th/9108028.



\bibitem{D'Hoker:1998}
  E.~D'Hoker and D.~Z.~Freedman,
  ``General scalar exchange in AdS(d+1),''
  Nucl.\ Phys.\  B {\bf 550}, 261 (1999)
  [arXiv:hep-th/9811257].



\bibitem{D.Freedman:1998}
  D.~Z.~Freedman, S.~D.~Mathur, A.~Matusis and L.~Rastelli,
  ``Correlation functions in the CFT($d$)/AdS($d+1$) correspondence,''
  Nucl.\ Phys.\  B {\bf 546}, 96 (1999)
  [arXiv:hep-th/9804058].

\bibitem{freedmanreview}
  E.~D'Hoker and D.~Z.~Freedman,
  ``Supersymmetric gauge theories and the AdS/CFT correspondence,''
  arXiv:hep-th/0201253.


\bibitem{W.Mueck:1998}
  W.~Mueck and K.~S.~Viswanathan,
  ``Conformal field theory correlators from classical scalar field theory  on
  AdS(d+1),''
  Phys.\ Rev.\  D {\bf 58}, 041901 (1998)
  [arXiv:hep-th/9804035].




\bibitem{Hong Liu:1998}
   H.~Liu, A.A.~Tseytlin,
  ``On Four-point functions in the CFT/AdS Correspondce,''
  arXiv:hep-th/9807097

\bibitem{I.S.Gradsbteyn}
   I. S. Gradshteyn, I. M. Ryshik:
Table of Integrals, Series, and Products


 
\end{thebibliography}
\end{document}